\documentclass[a4paper,11pt]{article}

\usepackage[]{geometry,color,graphicx}
\usepackage{array}
\begin{document}

\title{Fluorescence spectroscopy enhancement on photonic nanoantennas}

\author{J\'{e}r\^{o}me Wenger$^1$$^*$}

\date{}
\maketitle

$^{1}$Aix Marseille Univ, CNRS, Centrale Marseille, Institut Fresnel, UMR 7249, Marseille, France \\

$^*$ Corresponding author: jerome.wenger@fresnel.fr

\section{Introduction and motivation}%\label{sec1.1}

Despite the significant progress in single molecule fluorescence microscopy made over the last two decades, the efficient detection of a single molecule remains a major goal with applications in chemical, biochemical and biophysical analysis \cite{TinnefeldRev,PunjWires2014}. The phenomenon of light diffraction appears as a main physical limiting factor. Indeed, there is a huge size mismatch between a single molecule (below 5 nm) and the wavelength of light (around 500 nm). This size mismatch prevents the efficient interaction between an incoming light beam from a conventional optical microscope and a single fluorescent molecule, so the net detected fluorescence signal from a single molecule (ultimately defining the sensitivity and dynamic temporal resolution achievable) remain limited \cite{Novotnybook}.

Additionally, conventional optical microscopes are restricted to conditions of low density (or concentration) of fluorescent molecules \cite{TinnefeldRev,Levene}. In order to isolate a single molecule in the diffraction-limited volume of a confocal microscope, the concentration range must be typically in the pico to nanomolar range. However, a large majority of enzymes and proteins requires concentrations in the micro to millimolar range to reach relevant reaction kinetics and biochemical stability. Monitoring single molecules at high physiological concentrations thus requires overcoming the diffraction limit to confine light in a nanometer spot of volume in the zepto- ($10^{-21}$) or atto- ($10^{-18}$) liter range, more than three orders of magnitude below the femtoliter volumes achieved with confocal microscopes \cite{CraigheadRev,WengerRev}.

Confining light to the nanoscale can be achieved thanks to optical antennas \cite{Novotny2011}. Optical antennas are generally metallic nanostructures with dimensions much below the wavelength of light \cite{BrongersmaRev,HechtRev}. Like their radiofrequency counterparts, optical antennas convert propagating radiation into localized energy and vice-versa \cite{Novotny2011}. This opens new routes to enhance and control the emission from a single fluorescent emitter by improving the light-matter interaction between a single fluorescent molecule and the incoming beam, leading to the phenomenon of antenna-enhanced fluorescence emission \cite{Lakowicz09,HalasRev,Koenderink17}. Over thousand-fold enhancement of the single molecule fluorescence signal was reported with lithographically fabricated gold nanogap antennas~\cite{Kinkhabwala2009,Punj2013,Flauraud2017}, with dimers of gold nanoparticles assembled with DNA origami \cite{Acuna2012,Tinnefeld2015} and with a single gold nanorod \cite{Orrit13,Orrit14}.

%While metals provide nanoscale localization of intense light fields, resistive heating losses in metals can severely limit the performance of plasmonic antennas due to non-radiative energy transfer from the molecule to the metal. This fluorescence quenching phenomenon critically depends on the molecular distance to the metal as well as the fluorescence emission spectra position respective to the antenna mode resonances. Therefore, a balance must be found to maximize the net fluorescence enhancement.

In this Chapter, we will briefly introduce the key concepts to understand the phenomenon of fluorescence enhancement with optical nanostructures. The equations will remain simple, with the major goal to illustrate the different effects at play. Next we will provide an overview of different recent approaches to enhance single molecule fluorescence with plasmonic and non-plasmonic nanoantennas. Lastly, we will describe three main biochemical applications that this technique opens.

\section{Brief theoretical background: the physics of fluorescence enhancement}

In order to understand the phenomenon of enhanced fluorescence in the vicinity of a nanostructure, one has to go back to the representation of the quantum emitter as a two energy levels system. $\Gamma_{rad}$ and $\Gamma_{nr}$ are the rate constants for radiative and non-radiative transitions from the excited singlet state to the ground state. The total de-excitation rate from the excited singlet state is then $\Gamma_{tot} = \Gamma_{rad}+ \Gamma_{nr}$, and corresponds to the inverse of the excited state fluorescence lifetime $\tau = 1/\Gamma_{tot}$. With these notations, the ratio of the radiative rate to the total decay rate (probability of de-excitation transition accompanied with the emission of a photon) is defined as the fluorescence quantum yield $\phi = \Gamma_{rad}/\Gamma_{tot} = \Gamma_{rad}/(\Gamma_{rad}+\Gamma_{nr})$. Lastly, the excitation rate constant is noted $\sigma I_e$, with $\sigma$ being the excitation cross-section and $I_e$ the local excitation intensity.

Under steady-state conditions, the fluorescence brightness (or count rate) per molecule $Q$ can be written as \cite{Fluobouquin}
\begin{equation}\label{Eq:FM}
  Q =  \kappa \, \phi \, \frac{\sigma I_e}{1+I_e/I_s} \, .
\end{equation}
Here $\kappa$ denotes the light collection efficiency of the optical microscope apparatus, and $I_s = \Gamma_{tot}/\sigma$ is the so-called saturation intensity. Generally the experiments are performed in the weak excitation regime well below fluorescence saturation (corresponding to $I_e \ll I_s$), so as to avoid photobleaching the emitters and introducing other photodamages to the photonic nanostructure. In this regime, the brightness $Q$ reduces to a simpler form:\cite{LakowiczBook}
\begin{equation}\label{Eq:FMlow}
  Q =  \kappa \, \phi \, \sigma \, I_e \, .
\end{equation}
We retrieve here that the fluorescence rate per molecule is proportional to the collection efficiency, the quantum yield, and the excitation intensity.

In the presence of the optical nanoantenna, all the transition rates are potentially modified (here, we note with a $^{\ast}$ the quantities modified by the nanoantenna). The local excitation intensity becomes $I_{exc}^{\ast}$ and can be significantly greater than the incoming intensity by taking advantage of the subwavelength confinement of light at the nanoantenna (similar to the lightning rod effect) which can be further additionally enhanced by the plasmonic resonances of the nanoantenna. Locally, the excitation intensity $I_{exc}^{\ast}$ can thus overcome $I_{exc}$ by more than a hundred fold. The radiative decay rate is also enhanced (as a consequence of the reciprocity theorem) and becomes $\Gamma_{rad}^{\ast}$.

A major challenge in enhancing fluorescence with metal nanostructures comes from the presence of additional losses (with rate constant noted $\Gamma_{loss}^{\ast}$) which are introduced by the presence of the metal nearby. These losses account for non-radiative energy transfer to the free electron gas in the metal, whose energies are further dissipated by Joule (ohmic) losses. Consequently, the total decay rate with the nanostructure becomes $\Gamma_{tot}^{\ast} = \Gamma_{rad}^{\ast}+ \Gamma_{nr} + \Gamma_{loss}^{\ast}$. It is greater than the initial decay rate $\Gamma_{tot}$ so the net photodynamics decay is accelerated by the presence of the nanoantenna. However, this increase in the total decay rate $\Gamma_{tot}^{\ast}/\Gamma_{tot}$ must \textit{not} be confused with the fluorescence enhancement (which is the ratio of the brightnesses per emitter $Q^{\ast}/Q$). Indeed, a significant fraction of the total decay rate $\Gamma_{tot}^{\ast}$ can be due to non-radiative losses $\Gamma_{loss}^{\ast}$ which lead to a decrease of the apparent fluorescence brightness also known as quenching. Said differently, a fluorescent emitter can experience an ultrashort lifetime near a nanoantenna, corresponding to an increased total decay rate $\Gamma_{tot}^{\ast}$ but this can be essentially a consequence of strong non-radiative losses damping all the energy away from the radiative transitions. This situation is typically found for a fluorescent molecule in very close (below 5~nm) proximity to a small gold nanoparticle of diameter below 15~nm: the gold nanoparticle acts here as a fluorescence quencher. Therefore, in any fluorescence enhancement experiment with a nanoantenna, a delicate balance must be found between enhancing the radiative decay rate $\Gamma_{rad}^{\ast}$ without introducing too many losses $\Gamma_{loss}^{\ast}$. As a side note, we point out that the situation is conceptually different in surface enhanced Raman scattering (SERS), where the scattering is a nearly instantaneous process and where there is no finite lifetime of the excited state and no non-radiative decay rates involved. Note also that we assume here that the non-radiative decay rate $\Gamma_{nr}$ (intrinsic to the fluorescent emitter in use, set by its chemical structure) is not modified by the presence of the metal nanostructure ($\Gamma_{nr}^{\ast}=\Gamma_{nr}$). This assumption will simplify the coming equations and is still fully general as the additional losses effects are included in the term $\Gamma_{loss}^{\ast}$.

With the previous expression of the brightness per emitter $Q$, the fluorescence enhancement factor $\eta_F$ is then simply defined as the ratio of brightnesses with and without the photonic nanostructure. So the fluorescence enhancement factor writes
\begin{equation}\label{EqEta1}
  \eta_F = \frac{Q^{\ast}}{Q} = \frac{\kappa^{\ast}}{\kappa} \,\,\,\frac{\phi^{\ast}}{\phi} \,\,\,\frac{I_{exc}^{\ast}}{I_{exc}} \, .
\end{equation}
This equation indicates the different relevant quantities to enhance the net fluorescence signal per emitter. The collection efficiency $\kappa^{\ast}$ can be improved by the nanoantenna, directing more optical energy into the collection aperture of the microscope setup. The excitation intensity $I_{exc}^{\ast}$ can be locally enhanced and lastly the quantum yield $\phi^{\ast}$ can be modified.

Improving the apparent quantum yield of the fluorescent emitter deserves more comments to discuss the ratio $\phi^{\ast}/\phi$. First, one has to keep in mind that the quantum yield as a probability of radiative transition cannot exceed unity. Therefore, if the emitter has a high initial quantum yield $\phi$ without the nanostructure (close to 1, like Rhodamine6G molecules for instance), then there is not much that the optical nanoantenna can do to further increase this quantity, and the ratio $\phi^{\ast}/\phi$ will also be close to one (if it is not reduced to to quenching losses). On the contrary, if one starts with a low quantum yield emitter with $\phi$ on the order of a few percents, then the ratio $\phi^{\ast}/\phi$ can become apparently very large, thereby improving the net fluorescence enhancement factor $\eta_F$. Low quantum yield emitters experience thus the maximum benefit from the presence of the plasmonic nanoantenna. To better represent the influence of the quantum yield and the various transition rates, Eq.~(\ref{EqEta1}) can be rewritten
\begin{equation}\label{EqEta}
  \eta_F = \frac{\kappa^{\ast}}{\kappa} \,\,\, \frac{I_{exc}^{\ast}}{I_{exc}} \,\,\, \frac{\Gamma_{rad}^{\ast}}{\Gamma_{rad}} \,\,\, \frac{1}{1-\phi \,\, + \,\, \phi \, (\Gamma_{rad}^{\ast}+\Gamma_{loss}^{\ast})/\Gamma_{rad}} \, .
\end{equation}
Note that the last ratio in the denominator $(\Gamma_{rad}^{\ast}+\Gamma_{loss}^{\ast})/\Gamma_{rad}$ describes the influence of the nanoantenna that is only due to the photonic environment, including ohmic losses and compensating for the intrinsic chemical non-radiative decays $\Gamma_{nr}$. This ratio thus corresponds to the net increase in the local density of photonic states (LDOS), sometimes also referred to as the Purcell factor. A high LDOS may thus seem to lower the fluorescence enhancement $\eta_F$, but this is counter-balanced by the term $\Gamma_{rad}^{\ast}/\Gamma_{rad}$ in the numerator, so a high LDOS generally leads to a higher fluorescence enhancement.

Because of the dependence of the apparent value of the fluorescence enhancement $\eta_F$ on the initial quantum yield $\phi$, comparing between experimental results obtained with different molecular emitters must be made with caution. Extreme fluorescent enhancement factors above a thousand fold have been reported, but all use fluorescent dyes of quantum yield below 10\%, and a high fluorescence enhancement factor does not necessarily correlate with a bright photon count rate. To compensate for the dependency on $\phi$, the product $\eta_F \times \phi$ has been proposed as a figure of merit for the fluorescence enhancement \cite{Gill2012}. This solves one of the problem, but extra care still needs to be taken while comparing results to take into account the influence of other parameters such as the alteration of the collection efficiency and the saturation of the fluorescence process due to large excitation powers \cite{WengerIJO}.

\section{Experimental approaches to enhance fluorescence}

To overcome the diffraction limit, nanoantenna designs take advantage of sharp curvature radii, nanoscale gaps and plasmonic resonances, using metal nanostructures \cite{Novotny2011}. A large number of approaches have been investigated over the last decade in order to enhance the fluorescence emission of single molecules and quantum dots \cite{TinnefeldRev,PunjWires2014,Lakowicz09}. While classical top-down lithography like electron-beam lithography or focused ion beam milling remains the workhorse for fabricating the nanoantennas, alternative bottom-up strategies using nanoparticle assembly have recently seen much development driven by the seek for a simpler and higher throughput nanofabrication approach.

\begin{figure}[ht]
\begin{center}
\includegraphics[width=10cm]{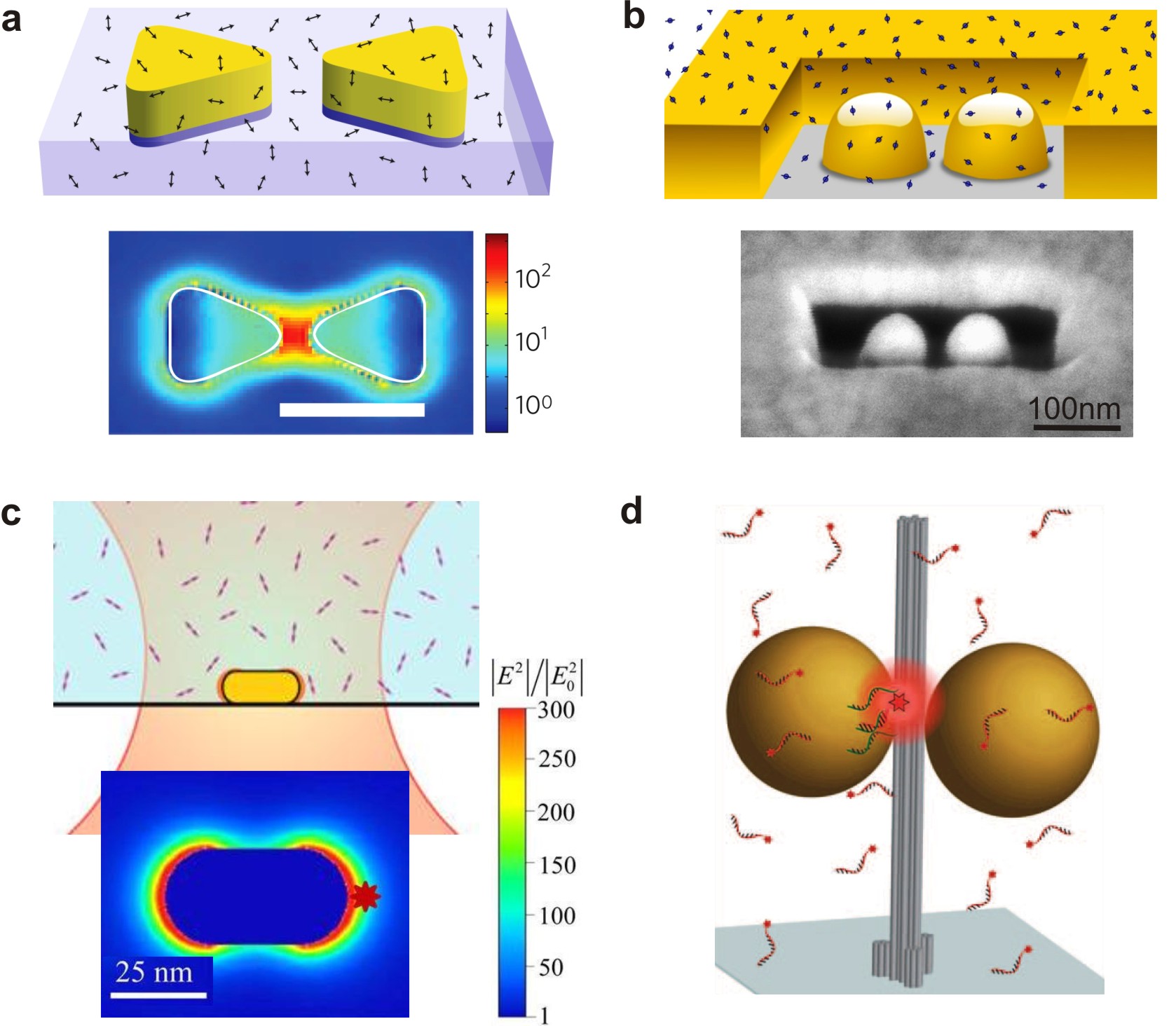}
\caption{Example of nanoantenna designs for single molecule fluorescence enhancement. (a) Bowtie nanoantenna covered by fluorescent molecules embedded in a PMMA resin \cite{Kinkhabwala2009}. The simulations represent the local excitation intensity enhancement around the nanoantenna (scale bar 100~nm). (b) Antenna-in-box designed for single-molecule analysis at high concentrations \cite{Punj2013}. (c) Single gold nanorod for enhanced single molecule detection in solution \cite{Orrit14}. The simulations show the near-field intensity map for a 47 x 25~nm gold nanorod. The excitation wavelength is 633 nm, and the excitation is polarized along the long axis of the nanorod. (d) Dimer nanogap antenna assembled by DNA origami \cite{Acuna2012}. Figures reproduced with permission: copyright (a) Nature Publishing Group 2009, (b) Nature Publishing Group 2013 (c) American Chemical Society 2014 (d) American Association for the Advancement of Science 2012.} \label{Fig:Ant}
\end{center}
\end{figure}

\subsection{Top-down milling}

Electron beam lithography, focused ion beam milling, or deep UV photolithography have a large flexibility in creating a variety of planar antenna designs at specific locations. A nice example is given by the bowtie antenna made of two facing nanotriangles with a 10~nm gap (Fig.~\ref{Fig:Ant}a) \cite{Kinkhabwala2009,MoernerFCS}. Plasmonic coupling between the nanoparticles leads to a local excitation intensity enhancement of 100-fold inside the nanogap region (so-called hot spot) where single fluorescent molecules are randomly deposited. By analyzing all the different single-molecule fluorescence traces on individual nanoantennas, fluorescence enhancement factors up to 1000-fold could be quantified, together with fluorescence lifetime down to 10~ps \cite{Kinkhabwala2009}.

In order to detect individual molecules diffusing in solution at high concentration, a special must be taken to ensure reducing the background fluorescence from molecules a few tens of nanometers away from the plasmonic antenna, yet still in the confocal detection volume. At the physiologically-relevant concentrations of a few micromolar, several thousands of molecules that are still present within the diffraction-limited confocal volume, and their non-enhanced emission can overwhelm the enhanced single-molecule signal from the nanoscale plasmonic hotspot. Specific strategies have been developed to address this issue, using low quantum yield emitters and/or adding a metal cladding layer. The design termed ``antenna-in-box'' combines a gap-antenna inside a nano-aperture, and has enabled single molecule detection at concentrations above 20~$\mu$M together with fluorescence enhancement above a thousand-fold (Fig.~\ref{Fig:Ant}b) \cite{Punj2013,Flauraud2017}.

Enhanced fluorescence spectroscopies applications need nanoantennas featuring a large-scale availability together with narrow gaps accessible to target the fluorescent probes. Recent advances in top-down lithography have been achieved towards this goal, using blurring-free stencil lithography patterning by dry etching through nanostencils \cite{Flauraud2015} or a combination of electron-beam lithography followed by planarization, etch back and template stripping \cite{Flauraud2017}.

\subsection{Bottom-up self assembly}

Bottom-up assembly of nanoparticles offers an attractive alternative to classical top-down nanofabrication thanks to its low operation cost, narrow gap sizes, use of single crystalline structures and potential large-scale availability \cite{LizMarzanRev,EdelRev}. A simple and direct approach relies on using a single nanoparticle of the shape of a sphere \cite{Estrada08,GuongFCS,GuongFCS2,LakowiczFCS,PunjOE13} or nanorod \cite{Orrit13,Orrit14,Orrit14PCCP}. Taking advantage of the local surface plasmon resonance enhancing and confining the electromagnetic field within a few nanometers near the nanorod apex, fluorescence enhancement above a thousand fold could be obtained with an antenna as simple as a single nanorod (Fig.~\ref{Fig:Ant}c) \cite{Orrit13,Orrit14}, enabling fluorescence correlation spectroscopy at high concentrations \cite{Orrit14PCCP}. Drying of nanoparticle colloidal suspension provides another straightforward access to resonant optical antennas with nanometer gap sizes \cite{Fan10}. For a dimer of 80~nm gold nanoparticles with 6~nm gap, 600-fold fluorescence enhancement and detection volumes down to 70~zL were achieved \cite{Punj15}. It is worth also mentioning that ultrafast emission of quantum dots down to picosecond lifetimes was achieved thanks to a patch antenna realized by a silver nanocube on a gold surface \cite{Akselrod14,Hoang15,Hoang16}.

To provide a better control on the nanogap distance and provide more reproducible antennas, the self-assembly of metal nanoparticles can be templated by DNA double strands \cite{Capasso11,Bidault11,Bidault12,Busson12,Bidault16} or DNA origami \cite{Acuna2012,Tinnefeld2015,Xu15,Acuna17}. This latter approach constitutes a powerful method to assemble nanoparticles into complex 3D designs with gap size tunability (Fig.~\ref{Fig:Ant}d). It also brings the key ability to bind the desired target molecule at the antenna hot spot \cite{Acuna2012}. Recent results use 100~nm diameter nanoparticles assembled on a DNA origami pillar with 12-17~nm gap featuring a binding site for a single fluorescent molecule. Fluorescence enhancement up to 5000-fold were reported, together with single-molecule detection at concentrations up to 25~$\mu$M \cite{Tinnefeld2015}.

\subsection{Dielectric alternatives to plasmonic metals}

The metals conventionally used for plasmonics (gold, silver, aluminum) still have significant absorption losses in the visible spectral range used for fluorescence spectroscopy applications. Energy damping to the free electron gas in the metal can severely quench the fluorescence emission, while simultaneously absorption of the laser beam induce Joule heating of the nanostructure \cite{Lakowicz09}. To find an alternative featuring lower absorption losses, all-dielectric nanoantennas based on silicon or germanium have been recently introduced \cite{GarciaEtxarri11,Evlyukhin12,Kuznetsov12,Albella2013}. Planar concentrators have been used to achieve extremely high collection efficiencies and detection rates of a single fluorescent molecule \cite{Lee11,Sandoghdar2014}. Silicon nanogap antennas (with a design very close to their gold counterpart) have been reported to enhance the fluorescence of a dye layer \cite{caldarola2015,Maiernew} and of single molecules \cite{Regmi2016Si}, with single molecule enhancement factors up to 270-fold. Although the gains reported remain lower than for the plasmonic metal counterparts, this approach is a promising alternative compatible with CMOS processing.

\begin{figure}[ht]
\begin{center}
\includegraphics[width=13cm]{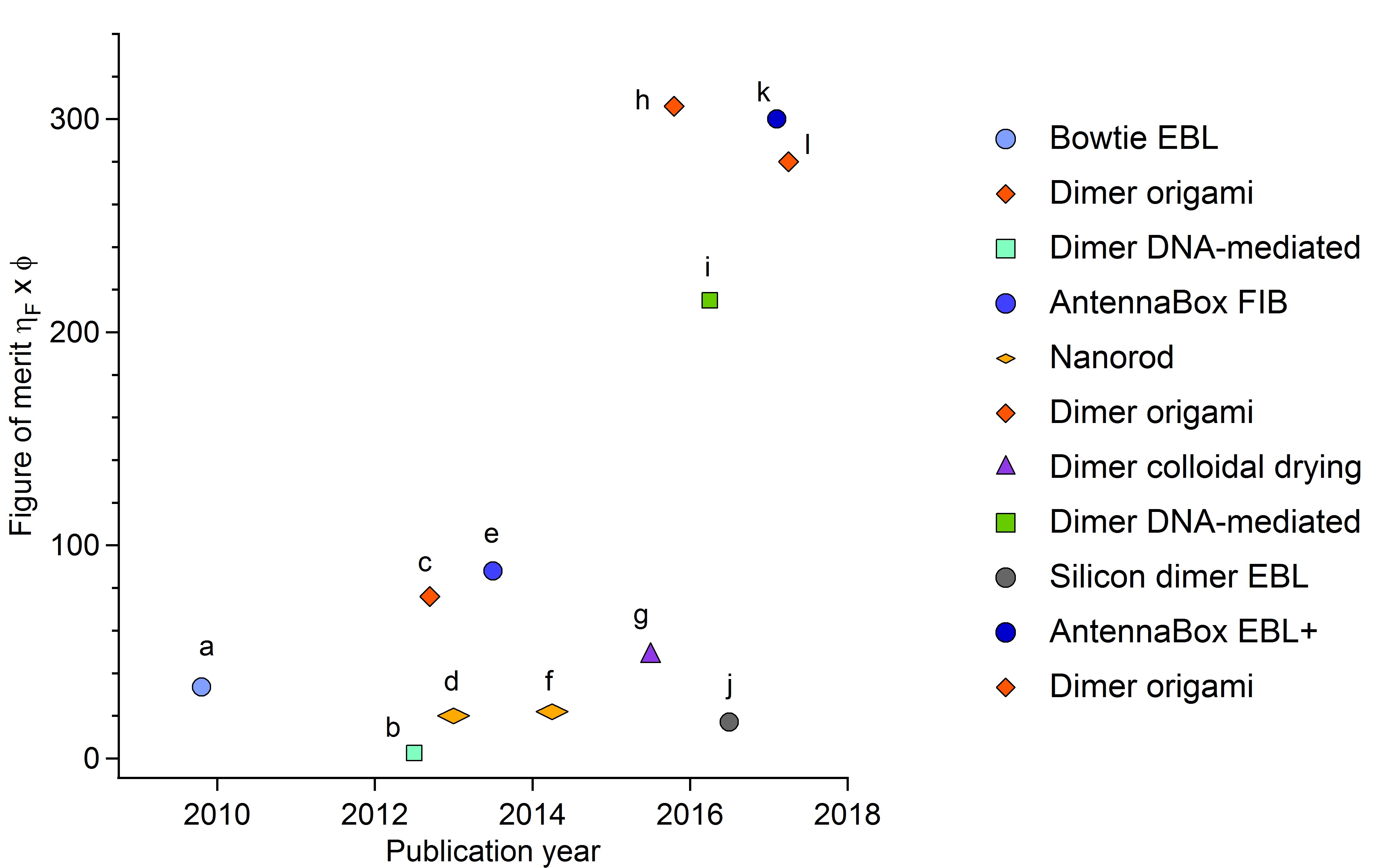}
\caption{Figure of merit for the single molecule fluorescence enhancement (defined as the product of the measured fluorescence enhancement factor $\eta_F$ times the initial quantum yield $\phi$ of the fluorescent emitter used) as a function of the publication year for different nanoantenna designs and fabrication approach. The letters written next to the data points relate to the following references: a \cite{Kinkhabwala2009}, b \cite{Bidault12}, c \cite{Acuna2012}, d \cite{Orrit13}, e \cite{Punj2013}, f \cite{Orrit14}, g \cite{Punj15}, h \cite{Tinnefeld2015}, i \cite{Bidault16}, j \cite{Regmi2016Si}, k \cite{Flauraud2017}, l \cite{Acuna17}.} \label{Fig:FoM}
\end{center}
\end{figure}

To conclude this section, Figure~\ref{Fig:FoM} compares the figure of merit for the single molecule fluorescence enhancement (defined as $\eta_F \times \phi$) for different techniques. As the techniques develop further and further, the figure of merit rises up to values around 300 which represent the current state-of-the-art at the end of 2017.

\section{Biochemical applications of enhanced fluorescence}

In this section, we discuss two driving applications for antenna-enhanced single molecule fluorescence detection: real-time DNA sequencing and living cell membrane investigations at the nanoscale. We also present the emerging field of single molecule F\"{o}rster resonance energy transfer (smFRET) enhanced by nanoantennas.

\subsection{Real-time DNA sequencing}

Achieving personalized quantitative genomics requires the development of novel methods for DNA and RNA sequencing that enable high-throughput, high-accuracy and low operating costs. Monitoring real-time single-molecule DNA and RNA sequencing by a single polymerase enzyme is a promising approach. So far, it has been achieved using circular nanoholes milled in an aluminum film (also called zero-mode waveguides \cite{Levene}) \cite{Eid09,Uemura10}. Each nanohole provides an observation chamber for watching the activity of a single polymerase enzyme at micromolar concentration of fluorescently-tagged oligonucleotides, each base (A,T,G,C) is color-coded with a spectrally distinct fluorophore. Every time a fluorescent nucleotide is incorporated into the DNA strand, there is a burst of fluorescence of millisecond duration which allows to retrieve the corresponding sequence as the replication is performed. A few thousand of nanoholes are currently monitored simultaneously, enabling massive parallelization \cite{Eid09}. So far, only circular aluminum nanoholes have been used for this application, but the sensitivity and detection rates could greatly benefit from the higher fluorescence enhancement offered by plasmonic nanoantennas.

\subsection{Nanoscale organization of lipid membranes}

Recent progress in cell biology indicates that the cell membrane features transient and fluctuating nanoscale assemblies of sterol and sphingolipids known as lipid rafts \cite{Sezgin17}. Investigating the role and formation of these nanodomains is of key interest for cell biology. However, the nanometer and microsecond resolutions that are simultaneously required fall far beyond the reach of standard microscopes \cite{Eggeling2009}. With their ability to confine light into nanometer dimensions  and drastically improve the fluorescence signal from a single molecule, optical nanoantennas offer a promising approach to investigate the nanoscale dynamic organization of living cell membranes \cite{Groves2012,Biteen2016,Orrit2016,Winkler2017,Regmi2017}. The flat geometry of the lipid bilayer membrane is remarkably well suited for the investigation by planar nanoantenna designs. Demonstrations on model lipid membranes \cite{Winkler2017} and CHO cell membranes \cite{Regmi2017} highlight the potential of nanoantennas to reveal nanodomains of 10~nm dimensions and sub-millisecond characteristic times. This fully bio-compatible approach opens interesting opportunities for living cell biophysics with single-molecule sensitivity at ultrahigh spatial and temporal resolutions.

\subsection{F\"{o}rster resonance energy transfer FRET}

As we have seen, optical nanoantennas can enhance the fluorescence signal from single quantum emitters. Interestingly, they can also be applied to enhance and modify the F\"{o}rster resonance energy transfer (FRET) between two nearby fluorescent emitters. FRET is one of the most widely used single molecule fluorescence techniques, with many applications to investigate the conformation dynamics of large molecules \cite{Weiss00} or perform biosensing \cite{Medintz03}. Metal nanoapertures \cite{Ghenuche14,deTorres15} and nanogap antennas \cite{Ghenuche15,BidaultFret16,deTorres16} have been shown to significantly enhance the FRET rate while simultaneously enabling single-molecule FRET at micromolar concentrations. Interestingly, the FRET gain is larger for more distant donor and acceptor molecules, which is interesting for the investigation of biochemical structures with donor-acceptor distances much beyond the classical F\"{o}rster radius \cite{Ghenuche14,Ghenuche15}. Moreover, the strongly inhomogeneous electromagnetic fields in plasmonic nanoantennas was also demonstrated to enhance FRET between nearly perpendicular donor and acceptor dipoles, enabling FRET detection that would otherwise be forbidden in a homogeneous confocal volume \cite{deTorres16}.

\section{Conclusion}

Optical nanoantennas overcome the classical diffraction limits of confocal microscopes and confine light down to dimensions close to the size of a single molecule. This enables to drastically improve the fluorescence emission rate, notably for low quantum yield emitters. Thanks to the recent advances in the nanofabrication, either top-down or bottom-up, the single molecule fluorescence toolbox is now significantly expanded. New directions involve single molecule fluorescence dynamics studies at physiologically-relevant micromolar concentrations and living cell membrane studies at ultrahigh spatiotemporal resolutions. The potential of nanoantennas is at hand to reveal new information on biological functions and dynamics.

%\section*{Acknowledgments}%\index{acknowledgments}
%RR is supported by the Erasmus Mundus Doctorate Program Europhotonics (Grant 159224-1-2009-1-FR-ERA MUNDUS-EMJD).

%\section*{Bibliography}%\index{bibliography}

%\bibliographystyle{psp-book-har}    % Bibliography: Author-Date system

\end{document}